\documentclass{ws-ijmpa}

\newcommand{\ltsimeq}{\raisebox{-0.6ex}{$\,\stackrel
           {\raisebox{-.2ex}{$\textstyle <$}}{\sim}\,$}}
\newcommand{\gtsimeq}{\raisebox{-0.6ex}{$\,\stackrel
           {\raisebox{-.2ex}{$\textstyle >$}}{\sim}\,$}}

\begin{document}

\markboth{D. M. Gingrich}
{Black Hole Production at the Large Hadron Collider}

\catchline{}{}{}{}{}

\title{BLACK HOLE PRODUCTION AT THE LARGE HADRON COLLIDER}

\author{DOUGLAS M. GINGRICH\footnote{Also at TRIUMF, Vancouver, BC V6T
2A3, Canada.}}

\address{Centre for Particle Physics, Department of Physics, University
of Alberta,\\
Edmonton, AB T6G 2G7, Canada\\  
gingrich@ualberta.ca}

\maketitle

\begin{history}
\received{\today}
\end{history}

\begin{abstract}
Black hole production at the Large Hadron Collider (LHC) is an
interesting consequence of TeV-scale gravity models.
The predicted values, or lower limits, for the fundamental Planck scale
and number of extra dimensions will depend directly on the accuracy of
the black hole production cross-section.    
We give a range of lower limits on the fundamental Planck scale that
could be obtained at LHC energies.   
In addition, we examine the effects of parton electric charge on black
hole production using the trapped-surface approach of general
relativity.  
Accounting for electric charge of the partons could reduce the black
hole cross-section by one to four orders of magnitude at the LHC.

\end{abstract}

\ccode{PACS numbers: 04.70.Bw, 04.50.+h, 12.60.-i, 04.70.-s}

\section{Introduction\label{sec1}}

Models of large\cite{Arkani98,Antoniadis,Arkani99} or
warped\cite{Randall99a,Randall99b} extra dimensions allow the
fundamental scale of gravity to be as low as the electroweak scale. 
For energies above the gravity scale, black holes can be produced in
particle collisions.
This opens up the possibility to produce black holes at the Large Hadron
Collider (LHC).
Once formed, the black hole will decay by emitting Hawking
radiation.\cite{Hawking75} 
The final fate of the black hole is unknown since quantum gravity will
become important as the black hole mass approaches the Planck scale.
If black holes are produced at the LHC, detecting them will not only
test general relativity and probe extra dimensions, but will also teach
us about quantum gravity.

Early discussions of black hole production in colliders postulated a
$\pi R_\mathrm{S}^2$ form for the cross-section, where $R_\mathrm{S}$ is 
the Schwarzschild radius of the black hole formed in the parton
scattering process.\cite{Banks,Dimopoulos01,Giddings01a}
Calculations based on classical general relativity have had limited
success in improving the cross-section
estimates.\cite{Eardley,Yoshino02c} 
The effects of mass, spin, charge, colour, and finite size of the
incoming particles are usually neglected.
The effects of finite size have been
examined\cite{Kohlprath,Giddings04} and only recently have
charge\cite{Yoshino06a} and spin\cite{Yoshino07} been
discussed.  
Attempts have been made to account for angular momentum in a heuristic
way by multiplying the simple expression for the cross-section by a form 
factor.\cite{Park01,Kotwal,Anchordoqui01,Ida02} 
Although these results are far from complete, they do indicated that the 
simple geometric cross-section is correct if multiplied by a formation
factor of order unity.\cite{Gingrich06a}

General relativistic calculations of the cross-section have usually 
been performed using the trapped-surface approach.
The two incoming partons are modelled as Aichelburg-Sexl shock
waves.\cite{Aichelburg}  
Spacetime is flat in all regions of space except at the shocks.
The union of these shock waves defines a closed trapped surface.
Black hole formation can be predicted by identifying a future trapped
surface, with no need to calculate the gravitational field. 

The trapped-surface approach was first applied to TeV-scale gravity
calculations by Eardley and Giddings\cite{Eardley} in four dimensions. 
Their work was extended to the $(n+4)$-dimensional case numerically by
Yoshino and Nambu.\cite{Yoshino02c} 
The numerical studies were improved by Yoshino and
Rychkov\cite{Yoshino05a} by analyzing the closed trapped surface on a 
future slice of spacetime.
These general relativistic calculations have enabled lower limits to be
obtained for the black hole mass produced by colliding particles in
TeV-scale gravity scenarios. 

Since black holes are highly massive objects, the momentum fraction of
the partons in the protons that form them must be high.
Thus typically valence quarks will be involved in black hole formation.
This means the most probable charge of the black hole in proton
collisions will be $+4/3$.
Since the gravitational field of each particle is determined by its
energy-momentum tensor, charge should affect the black hole formation.
First exploratory work by Yoshino and Mann\cite{Yoshino06a} obtained a
condition on the electric charges of the colliding particles for a
closed trapped surface to form.  
The results depend on the Standard Model brane
thickness.

In this paper, we take trapped energy into account and give limits on
the cross-section.
We derive lower limits on the Planck scale based on estimates
of the cross-section.\cite{Gingrich06a} 
We use the Yoshino and Mann charge condition in its
general form and build on their work by examining the effect of charge
on black hole production at the LHC.\cite{Gingrich06c} 

\section{Classical Parton Cross-Section\label{sec2}}

Black hole solutions in higher dimensions have a complicated dependence
on both the gravitational field of the brane and the geometry of the
extra dimensions.  
However, there are two useful approximations that may be made for a wide
class of models.
The brane on which the Standard Model particles live, is expected to
have a tension given by roughly the Planck or string scale. 
For black holes with mass $M$ substantially heavier than the fundamental
Planck scale in higher dimensions $M_D$, the brane's field should be
negligible and the production process for black holes should be
non-perturbative.     
We will assume that the only effect of the brane field is to bind the
black hole to the brane, and that otherwise the black hole may be
treated as an isolated object in the extra dimensions. 
If the geometrical scales of the extra dimensions $R$ (radii, curvature 
radii, variation scale of the warp factor) are all large compared to
$1/M_D$, there is a wide regime in which the geometry of the extra
dimensions plays no essential role.   
We consider black holes with horizon radius $r_\mathrm{h}$ much smaller
than the size of the extra dimensions, $1/M_D < r_\mathrm{h} \ll R$. 
Under these assumptions, it is often a good approximation to consider
the high-energy collision of the particles and the black hole formed to
be in $(n+4)$-dimensional flat spacetime.  

Since the black hole is not an ordinary particle of the Standard Model
and its correct quantum theoretical treatment is unknown, it is treated
as a quasi-stable state, which is produced and decays according to the
semiclassical formalism of black hole physics.
Using the above approximations, it has been argued
that at high energies black hole production has a good classical
description.\cite{Dimopoulos01,Giddings01a}  
This leads to the naive estimate that the cross-section for black hole
production is approximately given by the classical geometric
cross-section $\hat{\sigma} = \pi R_\mathrm{S}^2$, where $R_\mathrm{S}$
is the $(n+4)$-dimensional Schwarzschild radius corresponding to the
black hole mass. 
It depends on the fundamental Planck scale $M_D$ and the number of extra
dimensions $n$.
In the high-energy limit, the cross-section should depend on the impact
parameter $b$ between the two partons, and a range of black hole masses
will result for a given center of mass energy $\sqrt{\hat{s}}$. 
Since the cross-section is dominated geometrically by large impact
parameters $b \ltsimeq R_\mathrm{S}$, the average black hole mass should
be of the order of the center of mass energy, $\langle M\rangle \ltsimeq 
\sqrt{\hat{s}}$. 
It is often assumed that the black hole mass is given by $M =
\sqrt{\hat{s}}$.  

In studying the uncertainties in the classical parton cross-section, it
is useful to examine a more general form of the cross-section

\begin{equation} \label{eq1}
\hat{\sigma} = F \pi r_\mathrm{h}^2\Theta(M - M_\mathrm{min}) \, ,
\end{equation}

\noindent
where $F$ is a form factor (usually approximated as unity),
$r_\mathrm{h}$ is a more general horizon that may depend on the angular
momentum and charges of the black hole (usually taken to be the
non-rotating non-charged Schwarzschild radius in flat
$(n+4)$-dimensions), and $\Theta$ is a Heaviside step function that
allows black hole production only above some threshold mass
$M_\mathrm{min}$ (often implicitly assumed).   
In addition, $M < \sqrt{\hat{s}}$ needs to be considered to allow for
the possibility of not all the available energy being trapped behind the
horizon.  

There exists a threshold for black hole production.
In classical general relativity, two point-like particles in a head-on
collision with zero impact parameter will always form a black hole, no
matter how large or small their energy.
At small energies, we expect this to be impossible due to the smearing
of the wave function by the uncertainty relation.
This then results in a necessary minimal energy to allow the required
close approach.
The threshold is of order $M_D$, though the exact value is unknown since
quantum gravity effects should play an important role for the wave
function of the colliding particles.
For simplicity, it is usual to set this threshold equal to $M_D$.

In the high-energy limit, if the impact parameter is less than
$r_\mathrm{h}$, a black hole with mass $M \sim \sqrt{\hat{s}}$ can be
produced.  
To avoid quantum gravity effects and stay in the classical regime, we
require $M \ge M_\mathrm{min}$, where $M_\mathrm{min}$ should be a few
times larger than $M_D$, although it is often taken as $M_D$. 
A reasonable criterion for $M_\mathrm{min}$ is given by the requirement
of large entropy.\cite{Anchordoqui04}
In the following, we will find it useful to define the dimensionless
parameter

\begin{equation} \label{eq2}
x_\mathrm{min} = \frac{M_\mathrm{min}}{M_D} \, ,
\end{equation}

\noindent
and require $x_\mathrm{min} \gg 1$.

Throughout this paper, we use the Particle Data Group (PDG)\cite{PDG}
definition of the Planck scale

\begin{equation} \label{eq3}
M_D^{n+2} = \frac{1}{8\pi G_\mathrm{N}} \frac{1}{R^n} \, ,
\end{equation}

\noindent
where $G_\mathrm{N}$ is Newton's constant in four dimensions.

\section{Particle Cross-Section\label{sec3}}

Only a fraction of the total center of mass energy $\sqrt{s}$ in a
proton-proton collision is available in the parton-parton scattering
process. 
We define $s x_\mathrm{a} x_\mathrm{b} \equiv s \tau \equiv \hat{s}$,
where $x_\mathrm{a}$ and $x_\mathrm{b}$ are the fractional energies of
the two partons relative to the proton energies.
The full particle-level cross-section $\sigma$ is obtained from the
parton-level cross-section $\hat{\sigma}$ by using

\begin{equation} \label{eq4}
\sigma_{pp\to \mathrm{BH+X}}(s) = \sum_\mathrm{a,b} \int^1_\frac{M^2}{s}
d\tau \int^1_\tau \frac{dx}{x} f_\mathrm{a}\left(\frac{\tau}{x}\right)
f_\mathrm{b}(x) \; \hat{\sigma}_\mathrm{ab\to BH}(\hat{s}=M^2) \, ,
\end{equation}

\noindent
where $f_\mathrm{a}$ and $f_\mathrm{b}$ are parton distribution
functions (PDFs) for the proton. 
The sum is over all possible quark and gluon pairings.
Throughout this paper, we use the CTEQ6L1 (leading order with leading
order $\alpha_s$) parton distributions functions\cite{Pumplin} within
the Les Houches Accord PDF framework.\footnote{LHAPDF the Les Houches
Accord PDF Interface, Version 5.2.2, maintained by M. Whalley;
http://hepforge.cedar.ac.uk/lhapdf/.} 
We have taken $Q = R_\mathrm{S}^{-1}$ for the QCD scale.

Since $\hat{s} = M^2$, we can make a changing of variable from $\tau$ to
$M$ to obtain the differential cross-section in terms of parton
luminosity (or parton flux)

\begin{equation} \label{eq5}
\frac{d\sigma_{pp\to \mathrm{BH+X}}}{dM} = \frac{dL}{dM}\;
\hat{\sigma}_\mathrm{ab\to BH} \, , 
\ \mathrm{where}\quad
\frac{dL}{dM} = \frac{2M}{s} \sum_\mathrm{a,b} \int^1_{M^2/s}
\frac{dx}{x} f_\mathrm{a}\left( \frac{\tau}{x} \right) f_\mathrm{b}(x)
\, . 
\end{equation}

The differential cross-section thus factorizes for the case of $\hat{s}
= M^2$.
It can be written as the product of the parton cross-section time a
luminosity function. 
The parton cross-section is independent of the parton types and depends
only on the black hole mass, Planck scale, and number of extra
dimensions.
The parton luminosity function contains all the information about the
partons. 
Beside a dependence on black hole mass, it is independent of the
characteristics of the higher-dimensional space, i.e.\ the Planck scale
and number of extra dimensions.  
The particle-level cross-section does not truly factorize if the horizon
radius is used as the QCD scale in the parton density functions for the
proton. 

\section{Trapped Surfaces and Trapped Energy\label{sec4}}

Classical general relativistic calculations indicate that the mass of a
black hole formed in a head-on collision is somewhat less than the total
center of mass energy; the scattering is not completely inelastic.
Thus Eq.~(\ref{eq1}) should be modified by replacing the black hole mass
by a fraction of the available center of mass energy, leading to a
reduction in the cross-section.   

To improve the naive picture of colliding point particles, we need to
consider the grazing collision of particles in $(n+4)$-dimensional Einstein
gravity and investigate the formation of trapped surfaces.   
A common approach is to treat the creation of the horizon as a collision
of two shock fronts in Aichelburg-Sexl geometry.\cite{Aichelburg}
The Aichelburg-Sexl metric is obtained by boosting the Schwarzschild
metric to form two colliding shock fronts.
Due to the high velocity of the moving particles, spacetime before and
after the shocks is almost flat and the geometry can be examined for the
occurrence of trapped surfaces, which depend on the impact parameter.

Eardley and Giddings\cite{Eardley} developed a method for finding the
trapped surfaces for this system.
For a nonzero impact parameter, they were able to solve the problem
analytically for the $n=0$ case. 
They obtained limits on the final mass of the black hole formed and
found a range from $M > 0.71\sqrt{\hat{s}}$ for $b=0$ to $M >
0.45\sqrt{\hat{s}}$ for $b=b_\mathrm{max}$.  
This can be compared with a perturbative analysis that gave $M
\approx 0.8\sqrt{\hat{s}}$.\cite{Eath93}
For higher dimensions, they solved the $b=0$ case to obtain lower
bounds on the final black hole mass of $M > 0.71\sqrt{\hat{s}}$ to
$0.589\sqrt{\hat{s}}$ for $n=0$ to 7.   

Unfortunately the Eardley and Giddings results are not general enough to
be useful for nonzero impact parameters and higher dimensions, but they
indicate that a significant amount of the initial energy may not be
trapped behind the horizon.  
Understanding the case of a nonzero impact parameter in higher
dimensions is crucial to improving the cross-section estimates.
The analytic techniques used to study head-on collisions in general
relativity are not applicable to collisions at nonzero impact
parameter. 
Thus the claim that a black hole will be produced when $b <
r_\mathrm{h}$ can only be expected to be true up to a
numerical factor. 

Yoshino and Nambu\cite{Yoshino02c} solved this problem numerically for
$n>0$ and obtained the maximal impact parameter $b_\mathrm{max}$.
In their analysis, the trapped surface was constructed on the union of
the two incoming shocks. 
However, this slice of spacetime is not optimal in the sense that there
exists other slices located in the future.  
Yoshino and Rychkov\cite{Yoshino05a} improved the analysis by using a
future slice. 
They found that the lower bound on the black hole mass formed is never
more than 71\% of the available energy.
The fraction of energy available decreases with impact parameter and the
number of extra dimensions, from 0.71 to 0.46 for $n=0$ and from 0.59 to
0 for $n=7$. 
The mean lower bound on the trapped energies are about 0.6 and 0.27 for
$n=0$ and 7 respectively. 

The following describes one approach to taking estimates of the
non-trapped energy into account, and applying a minimum black hole mass
cutoff to final results.
We use the Yoshino and Rychkov\cite{Yoshino05a} data to obtain lower
bounds on the black hole cross-section.  

\section{Inelastic Particle Cross-Section\label{sec5}}

Previous calculations of the cross-section for producing a black hole
have neglected energy loss, and assumed that the mass of the created
black hole was identical to the incoming parton center of mass energy.
However, recent work\cite{Yoshino02c,Yoshino05a} shows the energy loss
to gravitational radiation is not negligible, and in fact is large for 
large number of extra dimensions and for large impact parameters.

The trapped mass $M$ is given by (using the notation of Anchordoqui
\textit{et al}.\cite{Anchordoqui03}) 

\begin{equation} \label{eq6}
M(z) = y(z) \sqrt{\hat{s}} \, ,
\end{equation}

\noindent
where the inelasticity $y$ is a function of $z \equiv b/b_\mathrm{max}$.
This complicates the parton model calculations, since the production of
a black hole of mass $M$ is lower than $\sqrt{\hat{s}}$ by $M/y(z)$, thus
requiring the lower cutoff on the parton momentum fraction to be a
function of the impact parameter.
We can no longer use the factorized version of the particle-level
cross-section given by Eq.~(\ref{eq5}). 

Following Anchordoqui \textit{et al}.\cite{Anchordoqui04,Anchordoqui03},
we take the proton-proton cross-section as the impact parameter-weighted
average over parton cross-sections, with the lower parton fractional
momentum cutoff determined by $M_\mathrm{min}$. 
This gives a lower bound $(x_\mathrm{min}M_D)^2 / (y^2s)$ on the parton
momentum fraction $x$. 
With this in mind, the $pp\to \mathrm{BH+X}$ cross-section becomes 

\begin{equation} \label{eq7}
\sigma_{pp\to \mathrm{BH+X}}(s,x_\mathrm{min}) \ge \int^1_0 2zdz
\sum_\mathrm{a,b} \int^1_{\frac{(x_\mathrm{min}M_D)^2}{y^2s}}
d\tau
\; \int^1_\tau \frac{dx}{x} f_\mathrm{a}\left( \frac{\tau}{x} \right)
f_\mathrm{b}(x)
\; \hat{\sigma}_\mathrm{ab\to BH}(\tau s) \, .
\end{equation}

\noindent
Since the amount of trapped energy is a lower bound, the resulting
cross-section is a lower bound.

\begin{figure}[tbp]
\begin{center}
\includegraphics[width=10cm]{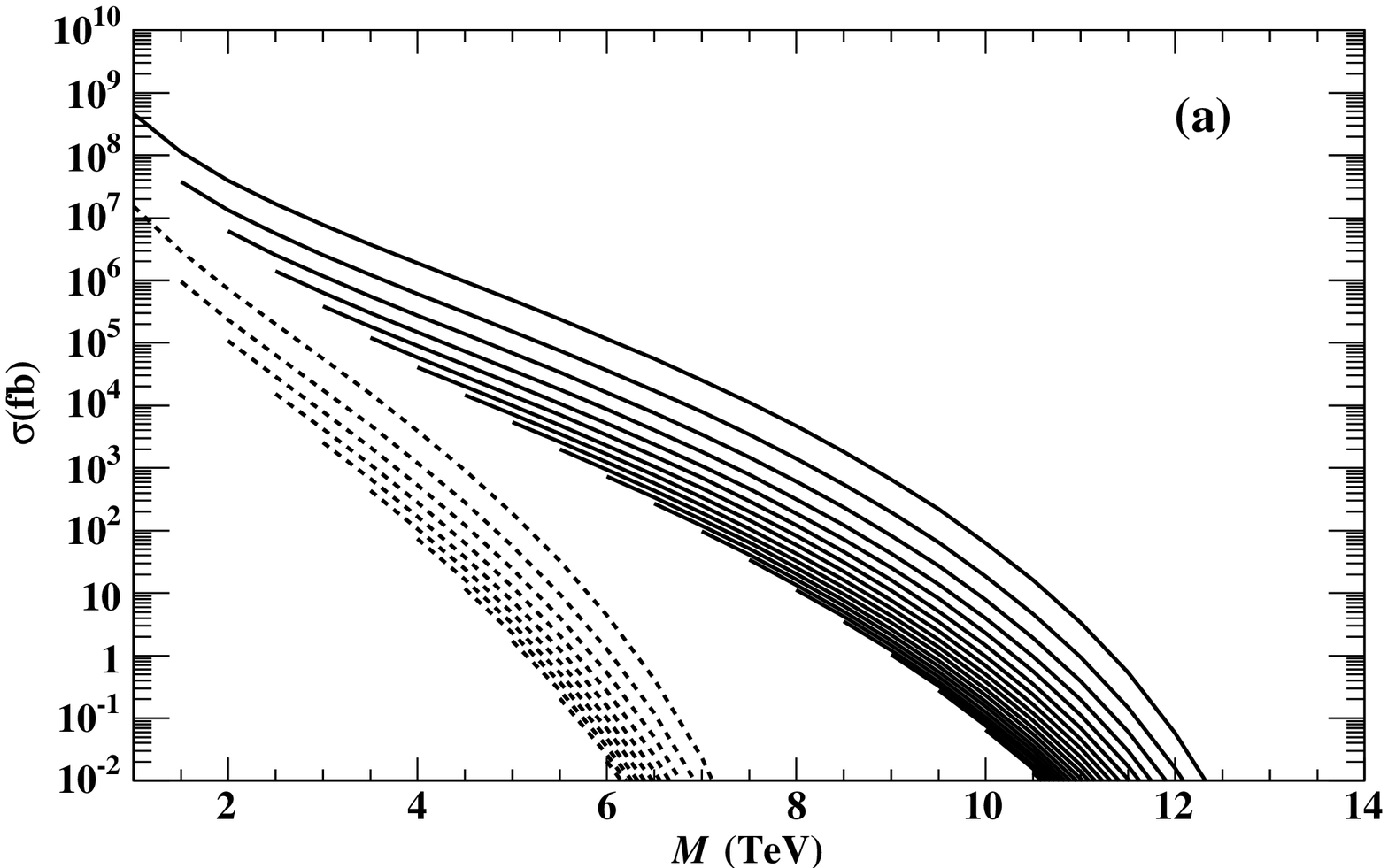}
\hfill
\includegraphics[width=10cm]{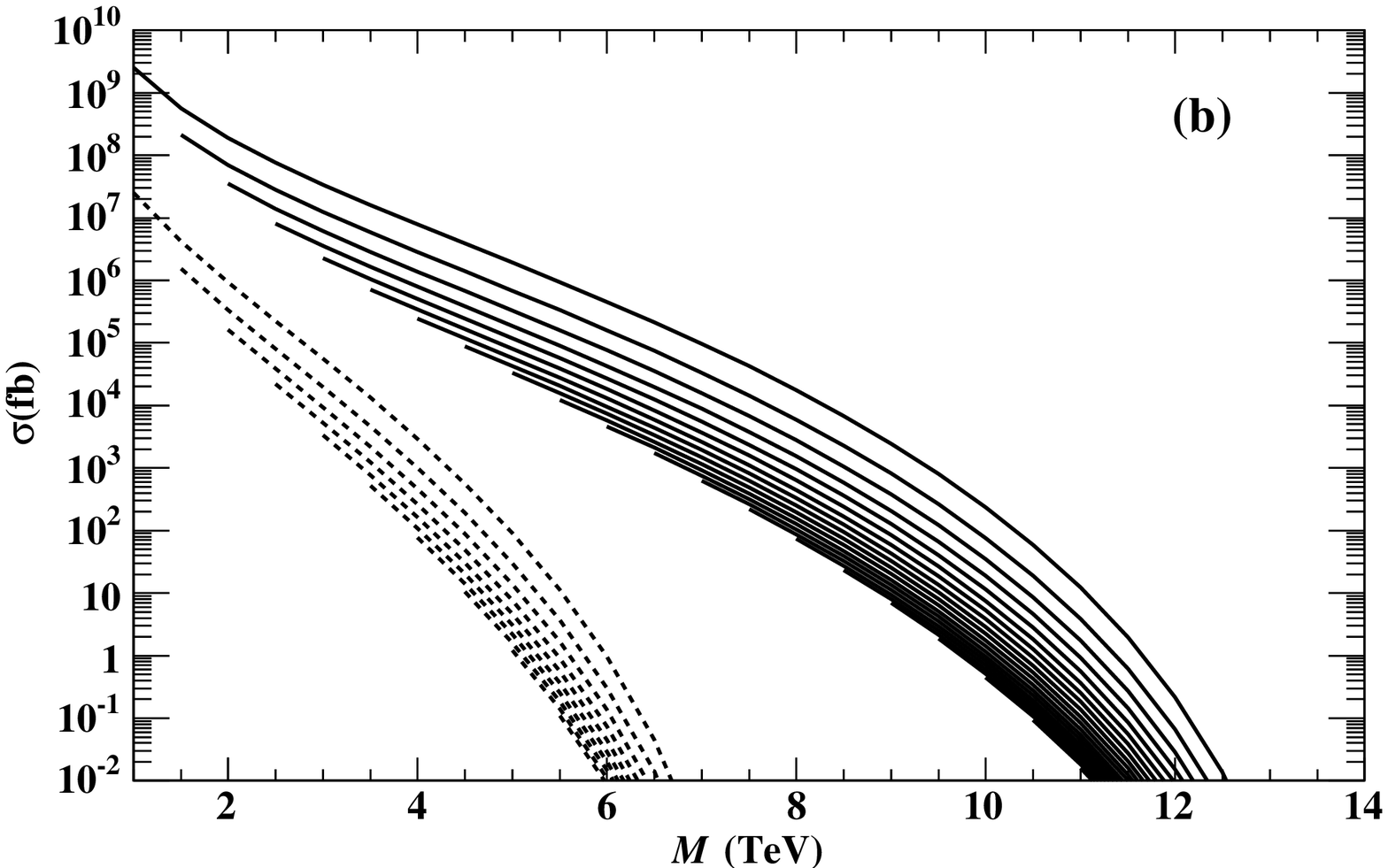}
\caption[]{Cross-section versus black hole mass. 
Solid curves classical cross-section and dashed curves trapped-surface
cross-section.    
Curves of same type for different Planck scales, 0.5~TeV top curves 
decreasing with increasing Planck scale in steps of 0.5~TeV. 
(a) $n=3$ and (b) $n=7$.
Ref.~\refcite{Gingrich06a}.\label{xsecpl}}
\end{center}
\end{figure}

\begin{figure}[tbp]
\begin{center}
\includegraphics[width=10cm]{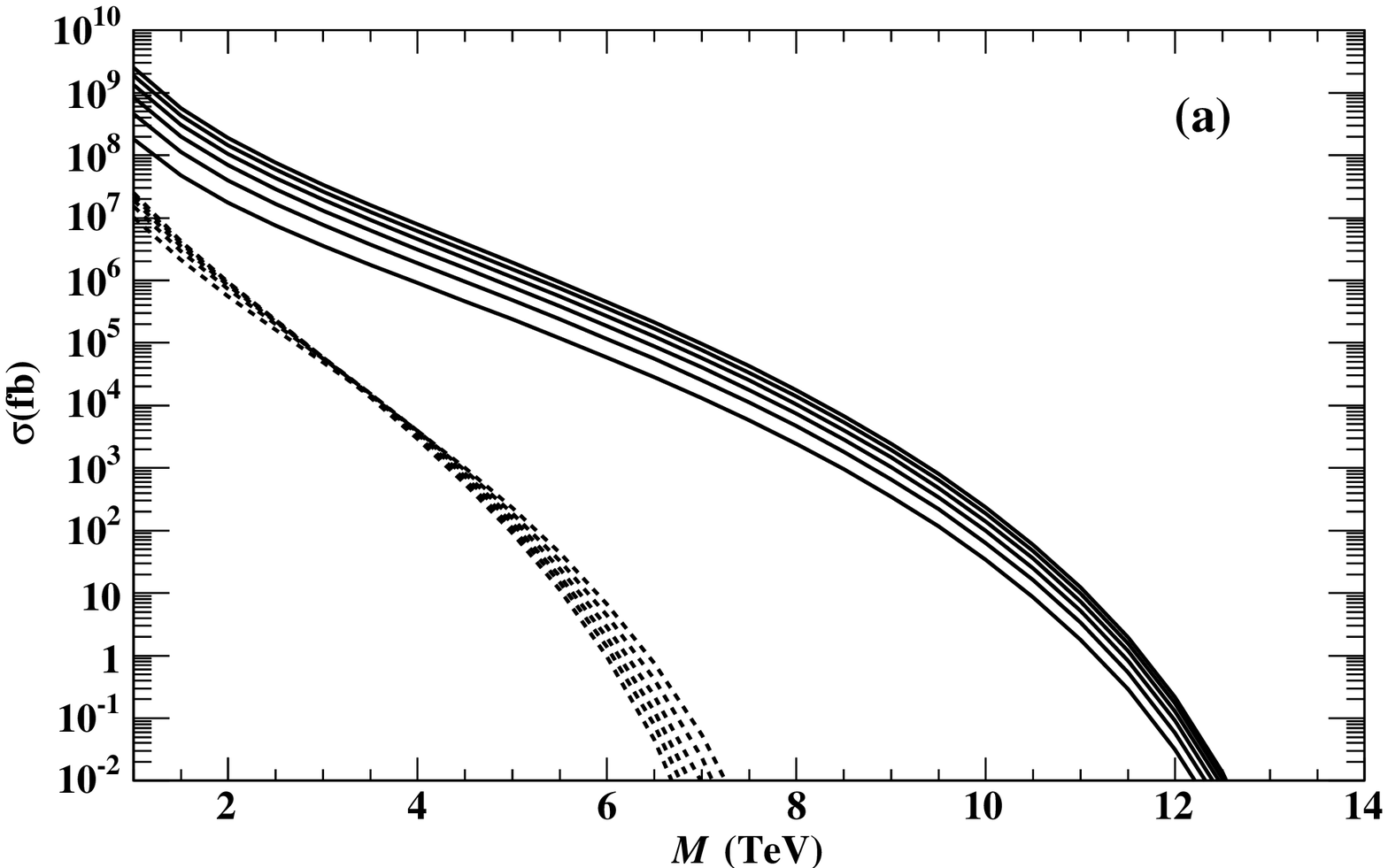}
\hfill
\includegraphics[width=10cm]{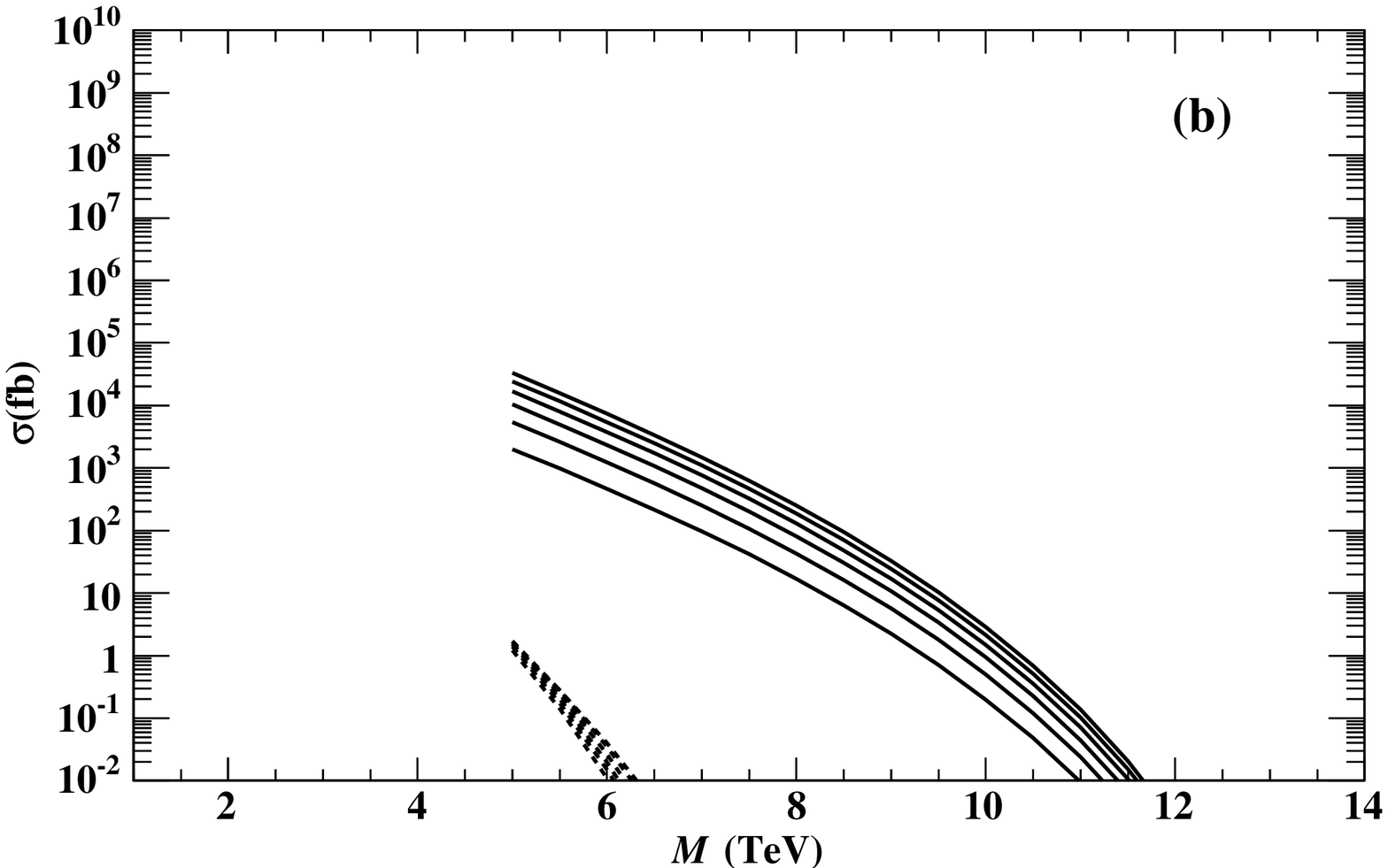}
\caption[]{Cross-section versus black hole mass.
Solid curves classical cross-section and dashed curves trapped-surface
cross-section.    
Curves of same type for different number of extra dimensions, top curves
$n=2$ and bottom curves $n=7$. 
(a) $M_D=1$~TeV and (b) $M_D=5$~TeV.
Ref.~\refcite{Gingrich06a}.\label{xsecn}}
\end{center}
\end{figure}

Taking $x_\mathrm{min} = 1$, we obtain the families of cross-section
curves shown in Figs.~\ref{xsecpl} and \ref{xsecn}.
The solid curves are for the classical cross-section calculated using
Eqs.~(\ref{eq1}) and (\ref{eq5}) with the form factors of
Ref.~\refcite{Yoshino05a}.   
We will henceforth refer to these curves as the classical cross-section.  
The dashed lower curves are given by Eqs.~(\ref{eq1}) and (\ref{eq7})
with the form factors of Ref.~\refcite{Yoshino05a}. 
We will henceforth refer to these curves as the trapped-surface 
cross-section.  
In Figs.~\ref{xsecpl}(a) and \ref{xsecpl}(b) the different curves of a given
type are for different Planck scales, starting from 0.5~TeV for the
top curve and decreasing with increasing Planck scale in steps of
0.5~TeV.
Figure~\ref{xsecpl}(a) is for $n=3$, while Fig.~\ref{xsecpl}(b) is for $n=7$. 
In Figs.~\ref{xsecn}(a) and \ref{xsecn}(b) the different curves of a given
type are for different numbers of extra dimensions, starting from
$n=2$ for the top curve and ending at $n=7$ for the bottom curve.
Figure~\ref{xsecn}(a) is for a Planck scale of 1~TeV, while
Fig.~\ref{xsecn}(b) is for a Planck scale of 5~TeV.  

The effect of non-trapped energy on the cross-section is large because
the LHC energy is close to the threshold for black hole production and
lost energy limits the availability energy for the black hole.
The cross-section curves show that there is less dependence on $n$ than
$M_D$.
This is because the $n$ dependence of the form factor tends to cancel
the $n$ dependence of the horizon radius.\cite{Webber}
It is reasonable to consider the classical cross-section with form
factors greater than unity as loose upper bounds on the black hole
cross-section, which may increase by a factor of a few as the formation
factors increase. 
We thus take the point of view that the black hole cross-section lies
between the classical and trapped-surface cross-sections.
The difference can be several orders of magnitude.
The trapped-surface cross-sections cut off at a mass above the
trapped-energy bounds.  
Applying a cutoff $x_\mathrm{min} > 1$ will further restrict the range
of the trapped-surface cross-sections, as well as the classical
cross-sections. 

\section{Lower Limits on the Planck Scale\label{sec6}}

The cross-sections in the previous section can be used to predict the
discovery limits for a given luminosity and be used, in principle, to
extract the Planck scale and number of extra dimensions.
In the event of no detectable black hole signal, the cross-sections can
also be used to set limits on the Planck scale and number of extra
dimensions.

\begin{figure}[htb]
\begin{center}
\includegraphics[width=10cm]{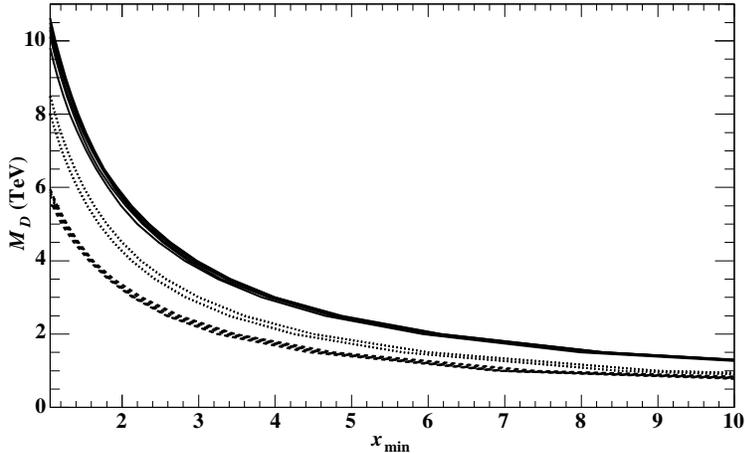}
\caption[]{Lower limits on Planck scale as function of cutoff
parameter.   
Solid curves classical case.
Dashed curves trapped-surface case.   
Dotted curves result of mass cutoff in trapped-surface case.
Spread in curves of same type due to different $n$.
Ref.~\refcite{Gingrich06a}.\label{limit}} 
\end{center}
\end{figure}

We consider the scenario in which no black hole signal has been observed
after the accumulation of an integrated luminosity of 300~fb$^{-1}$ at
$\sqrt{s} = 14$~TeV.
Rather than study the different decay phases of the black hole and
estimate the detector's capabilities for measuring them, we assume a
perfect detector.
This will give the most optimistic limits possible.
Assuming a perfect experiment, the 95\% confidence-level upper limit on
the cross-section is $10^{-2}$~fb.
Using this value of the cross-section, we have extracted lower limits on
the Planck scale $M_D$ as a function of cutoff parameter
$x_\mathrm{min}$ for different values of the number of extra dimensions
$n$. 
The results are shown in Fig.~\ref{limit} for $n = 2$ to 7.
The solid curves were obtained from the classical cross-sections.  
The dashed curves were obtained from the trapped-surface cross-section
bounds.  
The dotted curves are a result of the mass cutoff in the trapped-surface
cross-sections.  
The dotted curves can be consider as the infinite luminosity case of the 
trapped-surface predictions. 
The small spread in the different curves of a given type is due to
the different number of extra dimensions.

We can use Fig.~\ref{limit} to get a feel for how the different
cross-section models affect the range of Planck-scale limits.  
For $x_\mathrm{min} = 5$, a lower limit of $M_D > 2.4$~TeV is obtained for
the classical case and $M_D > 1.4$~TeV for the trapped-surface case.
The trapped-surface limit can be improved to $M_D > 1.7$~TeV with
infinite luminosity. 
Relaxing the cutoff criteria used to avoid quantum gravity effects to
$x_\mathrm{min} = 3$ gives a lower limit of $M_D > 3.8$~TeV for 
the classical case and $M_D > 2.2$~TeV for the trapped-surface case.
The trapped-surface limit can be improved to $M_D > 2.8$~TeV with
infinite luminosity. 
There appears to be very little sensitivity to the limits on the Planck
scale due to the number of extra dimensions: less than a 3\% effect.

\section{\boldmath{$(n+4)$}-Dimensional Reissener-Nordstr{\"o}m
Spacetime\label{sec7}} 

Yoshino and Mann\cite{Yoshino06a} have studied the effects particle
charge on black hole formation by replacing the usual flat
$(n+4)$-dimensional Schwarzschild metric by the flat $(n+4)$-dimensional
Ressener-Nordstr{\"o}m  metric, and searching for closed trapped
surfaces on the future lightcone.
They followed the usual Aichelburg-Sexl\cite{Aichelburg} prescription by
boosting a pair of metrics representing the gravitational fields of two 
point particles of mass $m$ and charge $q$ in $(n+4)$ dimensions in a
head-on collision to the likelike limit $(\gamma \to \infty)$.
In the process, the total energy $E = \gamma m$ and the quantity
$p_e^2 = \gamma q^2$ are keep fixed.
The charge dependence of the equivalent Achelburg-Sexl metric is
entirely contained in an additional term that depends on 

\begin{equation} \label{eq8}
a = \frac{2\pi(4\pi G_D p_e^2)}{n+1} \frac{(2n+3)!!}{(2n+4)!!} .
\end{equation}

\noindent
We refer to $a$ as a generalized charge.

The apparent horizon is given by the union of two surfaces that are a
monotonically increasing function $r$. 
When the two surfaces cross at the instance of collision
$r=r_\mathrm{min}$.
Imposing this boundary condition on the differential equation for the
apparent horizon gives\cite{Yoshino06a}

\begin{equation} \label{eq9}
x^4 = ( x - a_1 ) ( x - a_2 ) \, , 
\end{equation}

\noindent
where $x \equiv r_\mathrm{min}^{n+1}$, and $a_1$ and $a_2$ are given by
Eq.~(\ref{eq8}) with the corresponding values of $p_e^{(1)}$ and $p_e^{(2)}$.
Equation~(\ref{eq9}) determines the value of $r_\mathrm{min}$.
The apparent horizon exists if, and only if, there is a solution to
Eq.~(\ref{eq9}) with $x > a_1$ and $x > a_2$. 

\begin{figure}[htb]
\begin{center}
\includegraphics[width=7cm]{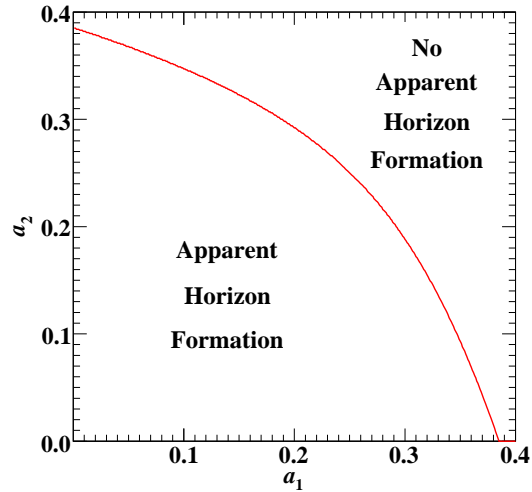}
\caption[]{Region of apparent horizon formation in
$(a_1,a_2)$-plane.
Ref.~\refcite{Gingrich06c}.\label{ah}} 
\end{center}
\end{figure}

Figure~\ref{ah} shows the region for apparent horizon formation in the
$(a_1,a_2)$-plane.
We see that both $a_1$ and $a_2$ must be sufficiently small for an
apparent horizon to form. 
For two particles of equal charge, $a_1 = a_2 = 1/4$ is a solution of
Eq.~(\ref{eq9}).  
For one charged particle and one neutral particle, $a_1 = 2/(3\sqrt{3})$
and $a_2 = 0$ is also a solution of Eq.~(\ref{eq9}).   

We can understand the requirement on $a_1$ and $a_2$ physically as
follows. 
Since $a_1$ and $a_2$ are proportional to $(p_e^{(1)})^2$ and
$(p_e^{(2)})^2$, the condition derived in Eq.~(\ref{eq9}) does not
depend on the sign of the charge of either particle.
This is because the gravitational field due to each charge is generated
by an electromagnetic energy-momentum tensor
$T_{\mu\nu}^{(\mathrm{em})}$ that depends on the square of the charge. 
The gravitational field induced by  $T_{\mu\nu}^{(\mathrm{em})}$ of the
incoming particles is repulsive, and its affect becomes dominant around
the center. 
As the value of $a$ increases, the repulsive region becomes larger,
preventing formation of the apparent horizon.
The critical value of $a$ for apparent horizon formation occurs when the 
repulsive gravitational force due to the electric field becomes
equivalent to the self-attractive force due to the energy of the system.

The approach for handling the confinement of the electric field
to the Standard Model three-brane is far from clear.
So far, we have ignored this effect by using the $(n+4)$-dimensional
Einstein-Maxwell theory.
We develop the relationship between the electric charge in four
dimensions $q_4$ and the charge in higher-dimensional Maxwell theory
$q$.    
For two particles in $(n+4)$ dimensions with the same charge at rest, for
example, the force between them is $F = q^2/r^{n+2}$.
If the gauge fields are confined to the Standard Model brane, the only
characteristic length scale is the width of the brane, which should be
of the order of the Planck length.
We introduce the constant $C_\mathrm{brane}$: $1/M_D \to
C_\mathrm{brane}/M_D$, where $C_\mathrm{brane}$ is a dimensionless
quantity of order unity.  
For sufficiently large $r$, 

\begin{equation} \label{eq10}
q^2 = q_4^2 \left( \frac{C_\mathrm{brane}}{M_D} \right)^n \, .
\end{equation}

\noindent
The brane thickness is a measure of how confined the Standard Model
electric charge is to the brane.
If the brane is thick, the Maxwell theory would be higher dimensional in
the neighbourhood of the particle.
We let $q_4^2 = C_q^2 \alpha$, where $C_q$ is the charge in units of
elementary charge $e$ ($-1/3$ or $+2/3$ for quarks and 0 for gluons) and
$\alpha$ is the fine structure constant.  
Our treatment of the electric charge has not fully taken into account
the effects of confinement of the electric field on the brane.
We have also ignored the brane tension and the structure of the extra
dimensions. 

From the above, the general charge becomes

\begin{equation} \label{eq11}
\frac{a}{r_0^{2(n+1)}} = C_q^2 \alpha \left( \frac{M_D}{m} \right)
\left( \frac{M_D}{E} \right) \pi \frac{\Omega_{n+1}^2}{n+1}
\frac{(2n+3)!!}{(2n+4)!!} \left( \frac{C_\mathrm{brane}}{2\pi} \right)^n
\, ,
\end{equation}

\noindent
where $r_0 = \left( 8\pi G_D E/\Omega_{n+1} \right)^\frac{1}{n+1}$.

Choosing values for $C_q$ and $m$, we can use Eq.~(\ref{eq11}) to
study the condition for apparent horizon formation as a function of $n$,
$M_D$, and $C_\mathrm{brane}$.   
An apparent horizon will not occur at the instance of collision if the
brane is thick or if the spacetime dimensionality is low. 
Charge effects will not be significant at high energies.

\section{Effect of Charged Partons on the Cross-Section\label{sec8}}

We work with parton luminosity, which is independent of $n$ and $M_D$.
Only the condition on which quarks to include in the sum of
Eq.~(\ref{eq5}) depends on $n$ and $M_D$.
Thus the upper and lower bounds on the parton luminosity do not change
for different parameters. 
To a good approximation, we can ignore the contribution from the sea
quarks at high black hole masses. 
The gluon contribution is the lower bound on the luminosity function
when the charged quarks do not contribute to the cross-section.   
We take the running of the coupling constant into account; we choose
$\alpha$ equal to 1/124 in the following calculations. 
Because of the large momentum transfer in black hole production, we use
current quark masses. 
Quark masses of $m_\mathrm{d} =8$~MeV and $m_\mathrm{u} =4$~MeV are
chosen for the valence quarks in the proton.
To study Eq.~(\ref{eq5}), we must first calculate Eq.~(\ref{eq11}), and
then determine if the condition in Fig.~\ref{ah} is satisfied. 
If it is, the parton pair is included in the sum in Eq.~(\ref{eq5}).

\begin{figure}[htb]
\begin{center}
\includegraphics[width=10cm]{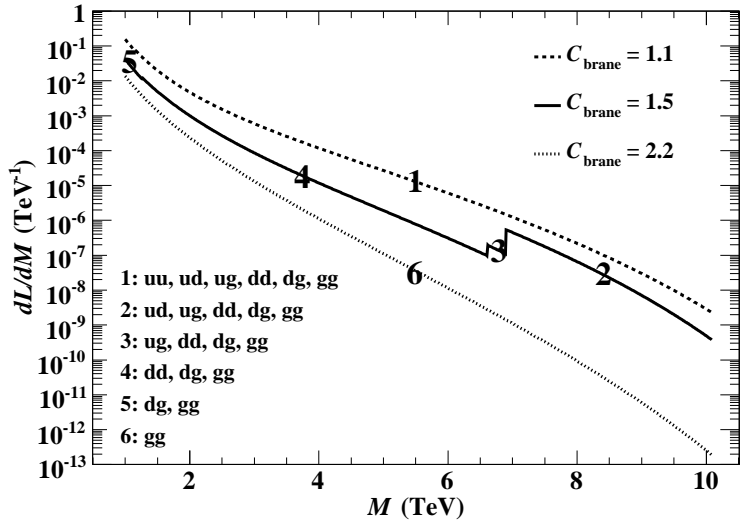}
\caption[]{Parton luminosity function versus black hole mass with 
charge condition applied for different brane thicknesses for $n=7$
and $M_D=1$~TeV.  
Numbers on plot show different parton contributions to parton luminosity
function. 
Ref.~\refcite{Gingrich06c}.\label{charge}} 
\end{center}
\end{figure}

Figure~\ref{charge} shows the parton luminosity for different brane
thicknesses for seven extra dimensions and a Planck scale of 1~TeV. 
The top curve is the case when all the partons contribute to the
cross-section, while the lower curve is the case when only the neutral
gluons contribute to the cross-section. 
The contributions of the different quarks in the intermediate region
depends on $M$, $n$, $M_D$, and $C_\mathrm{brane}$.
The thresholds for different quarks to contribute occur as a function of 
$M$ for fixed $n$, $M_D$, and $C_\mathrm{brane}$. 
The location of the thresholds may or may not occur in the mass region
of our plot. 
From Fig.~\ref{charge}, we see that charge can affect any black hole mass
and the effect is very sensitive to the brane thickness.
The decrease in parton luminosity, and thus cross-section, can range from
about one to four orders of magnitude over a black hole mass range of
1 to 10~TeV due to charge effects. 
The cross-section is nontrivial only over a range of brane thicknesses
from 1.1 to 2.2.
Plots with different number of dimensions are similar to
Fig.~\ref{charge}; they are always bounded above and below by the same
values, but for different values of the brane thickness.

\begin{figure}[htb]
\begin{center}
\includegraphics[width=10cm]{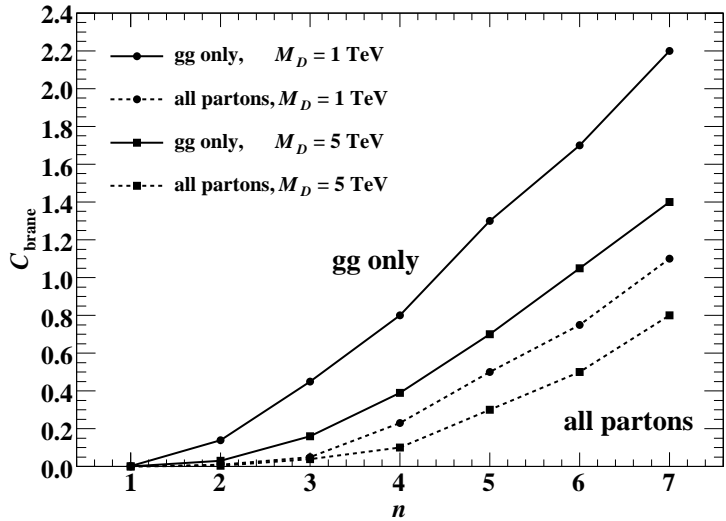}
\caption[]{Brane thickness versus number of dimensions for which all quarks
(dashed lines) and no quarks (solid lines) contribute to cross-section.   
Circles are for $M_D=1$~TeV and squares are for $M_D=5$~TeV.  
No quarks contribute to cross-section in region above solid curves.  
Cross-section is not affect by charge condition below region of dashed
curves.  
Some quarks contribute to cross-section in region between different
curve types. 
Ref.~\refcite{Gingrich06c}.\label{brane}}
\end{center}
\end{figure}

For each number of dimensions, we determine the maximum brane
thickness for all partons to be included in the parton luminosity and
the minimum brane thickness for only gluons to be included in the parton
luminosity. 
The results are shown in Fig.~\ref{brane}.
For a thin brane, the cross-section is not affected for high dimensions.  
For a thick brane, the cross-section is reduced for most number of dimensions.
For a Planck scale of 1~TeV and a brane thickness of 1~TeV$^{-1}$, the
cross-section is minimal for $n\ltsimeq 4$, not affected for $n\gtsimeq
7$, and has a range of values in the region $5\ltsimeq n\ltsimeq 6$. 

\section{Discussion\label{sec9}}

The limits on $M_D$ presented here are compatible with the discovery
limits that have been determined in previous
work.\cite{Anchordoqui04,Anchordoqui03,Giudice01,Ahn05,Tanaka,Harris05a} 
Our limits might appear different due to the stringent requirements on
$x_\mathrm{min}$ and the different definition of $M_D$.
The large difference between the classical and trapped-surface
cross-sections does not translate into a large difference in the limits
on $M_D$ because both cross-sections fall rapidly at low values of the
cross-section.  
If one is willing to relax the requirement on $x_\mathrm{min}$ and risk
entering the quantum-gravity regime, than the differences between the two
models becomes significant.
This difference presumably still holds when the uncertainties in the
black hole decay and experimental effects are taken into account.

The trapped-energy approach only gives a lower bound on the final mass 
of the black hole.  
In order to clarify the final mass, different methods such as the direct
study of gravitational wave emission are necessary.\cite{Cardoso05}  
The problem is extremely difficult because of the nonlinearity of
Einstein's equations, and because the high-energy collision of the two 
particles producing a black hole requires inclusion of nonlinear effects.

Taking the boosted Reissner-Nordstr{\"o}m metric as a reasonable
description of ultarelativistic quarks, we have shown that charge
effects will significantly decrease the rate of black hole formation at
the LHC, if the brane is somewhat thick or if the number of extra
dimensions $n$ is not too large. 
The charge effects can be quite large because the electromagnetic
energy-momentum tensor is proportional to $p_e^2 \sim \gamma\alpha$ and
the Lorentz factor $\gamma$ is much larger than $1/\alpha$ for
ultrarelativistic quarks.

There are many uncertainties in our understanding of black hole
production in higher dimensional TeV-scale gravity. 
Reliable predictions of the cross-section are not yet available.
We have explored some options for filling in the gaps in our
understanding. 
In this way, we hope to be better prepared to confront the possibility
of black hole production at the LHC.


\end{document}